\documentclass[aps,prb,showpacs]{revtex4}

\usepackage{graphicx}
\usepackage{amssymb}

\begin{document}

\title{Half-Metallicity of LSMO}

\author{G. Banach$^{1,2}$ and W.M. Temmerman${^1}$}
\affiliation{
$^{1}$ Daresbury Laboratory, Daresbury, Warrington WA4 4AD, UK \\
$^{2}$ Institute of Low Temperature and Structure Research, 
     Polish Academy of  Science, Wroclaw , Poland}

%\corauth[coresp]{Tel.: +44 1925 603153; fax: +44 1925 603172;\ead{G.Banach@dl.ac.uk}}

\date{\today}

\begin{abstract}
Self-interaction corrected local spin density approximation calculations were performed for 
La$_{(1-x)}$Sr$_x$MnO$_3$ (LSMO) ($0.0<x<0.5$). 
The influence and inter-relationship of Sr doping, magnetic structure, O displacements and 
phase segregation on the Mn charge state were studied. 
A half-metallic state was obtained for LSMO with manganese configuration Mn$^{3+}$, 
whilst Mn$^{4+}$ gave rise to a metallic state with a negligible spin polarisation at the Fermi level.
Elongating the MnO$_6$ octahedron led to a static mixed valence Mn$^{3+}$/Mn$^{4+}$ configuration.
In the mixed valence state the total energy was minimized by an ordered array
of Mn$^{4+}$ and Mn$^{3+}$ MnO$_2$ planes which showed charge ordered stripes. 
\end{abstract}

%\keywords{Half metals, LSMO}


\maketitle

% main text

\section{Introduction}
The half-metallic properties of La$_{(1-x)}$Sr$_{x}$MnO$_{3}$ (x=0.3) (LSMO) 
are of great importance for applications in spintronics. The 
tunnel magnetoresistance junction of LSMO/SrTiO$_3$/LSMO 
shows magnetoresistance ratio in excess of 1800\% \cite{bowen}. 
This result, according to Ref. \onlinecite{bowen}, strongly underlines the half-metallic
nature of mixed-valence manganites.
The electronic properties of LSMO, as described by band 
theory, are nearly half-metallic \cite{picket97,picket98,picketJAP,livesay}, 
reflecting the so-called transport half-metallic behaviour.\cite{nadgorny,mazin} 
However the 
fascinating electronic and magnetic properties of LSMO, 
including colossal magnetoresistance (CMR), indicate that 
the electronic structure is more complex than the standard 
band theory picture (see reviews \cite{coeyrev,tokura}).
In particular, the electronic 
structure is determined by the competition of double 
exchange and superexchange interactions, charge/orbital 
ordering instabilities, and strong coupling to the lattice 
deformations. Local Jahn-Teller effects, such as random Jahn-Teller distortions of the MnO$_6$ \cite{dzero} octahedra,
as well as dynamical effects \cite{michaelis}, have recently been invoked to explain 
the magnetoconductivity and optical conductivity respectively. Despite the numerous studies of the 
phase diagram of LaMnO$_3$-SrMnO$_3$ (LMO-SMO) 
(such as in Refs. \onlinecite{hemberger, maezono, majewski}),
there are still many conflicting interpretations of the role of the Jahn-Teller effect
in this material \cite{despina,cohn}, localisation of {\it d} electrons, \cite{saitoh,subias} and 
polarisation of electrons at the Fermi level.\cite{park}

In this paper we discuss issues concerning the charge 
ordering, and more specifically, the distribution of Mn$^{3+}$ and 
Mn$^{4+}$ in LSMO. We use the first principles self-interaction 
corrected local spin density (SIC-LSD) approximation.\cite{sic} 
This method can determine the number of valence band 
states. Hence, it can differentiate between Mn contributing three 
states (Mn$^{3+}$) to the valence band with the remaining four Mn 
states (3$t_{2g}$ and 1$e_{g}$) localised well below the valence band, 
and Mn contributing four states (Mn$^{4+}$) to the valence band, with 
the remaining three Mn states (3$t_{2g}$) localised below the 
valence band. This method was successfully applied to the 
study of orbital order in LaMnO$_3$.\cite{rik}

Calculations with the SIC-LSD as a function of Sr doping are presented 
in this paper for La$_{1-x}$Sr$_{x}$MnO$_{3}$ where 0.0$\le$x$\le$0.5.
The Sr doping is modeled with supercells and also with the rigid band approach.
In particular we will be concerned with determining the Mn valency as a function of concentration x.
It is found, as expected, that the valency changes as a function of Sr doping from
Mn$^{3+}$ to Mn$^{4+}$. However, a mixed phase of Mn$^{3+}$ 
and Mn$^{4+}$ valencies is found to accompany the valency change. 
The Mn$^{3+}$/Mn$^{4+}$ ordering in this mixed phase is consistent 
with charge ordered stripes. Furthermore it is found that, in the Mn$^{4+}$
state and as a function of Sr doping, a change of magnetic structure from
ferromagnetic to anti-ferromagnetic takes place. 
To highlight the effects of Sr doping on the Mn valencies the
lattice parameters are kept constant for all concentrations of Sr doping.
However, a slightly larger lattice parameter makes the Mn$^{3+}$ valency
less unfavourable. The structural properties of LSMO show an
equally rich variety of charge ordered states. Elongating
the MnO$_{6}$ octahedron makes it for the Mn ion more likely
to take on the Mn$^{3+}$ valency.   

The paper is organized as follows. In the next section
we introduce the theoretical background of our
electronic structure calculations, and in particular,
the self-interaction corrected local spin density method.
The technical and computational details regarding the
application of the SIC-LSD to LSMO are discussed in Section III.
In Section IV, we verify that the method works separately for LaMnO$_3$ and SrMnO$_3$.
Section V discusses the correlation between magnetic structure and charge order in LSMO.
The influence of oxygen displacements on the charge order is investigated in Section VI.
Section VII presents the results of calculations for various realizations of 
phase separation and their influence on the charge order. 
The conclusions of the paper are summarized in Section VIII.

\section{Theory}

The basis of the SIC-LSD formalism is a self-interaction free total energy functional, 
\( E^{SIC} \), obtained by subtracting from the LSD total energy functional,
\( E^{LSD} \), a spurious self-interaction of each occupied electron state 
\( \psi _{\alpha } \)\cite{pedrew}, namely 
\begin{equation}
\label{eq1}
E^{SIC}=E^{LSD}-\sum _{\alpha }^{occ.}\delta _{\alpha }^{SIC}.
\end{equation}
 Here \( \alpha  \) numbers the occupied states and the self-interaction correction
for the state \( \alpha  \) is 
\begin{equation}
\delta _{\alpha }^{SIC}=U[n_{\alpha }]+E_{xc}^{LSD}[\bar{n}_{\alpha }],
\end{equation}
with \( U[n_{\alpha }] \) being the Hartree energy and \( E_{xc}^{LSD}[\bar{n}_{\alpha }] \)
the LSD exchange-correlation energy for the corresponding charge density \( n_{\alpha } \)
and spin density \( \bar{n}_{\alpha } \). 
The SIC-LSD approach can be viewed as an extension of LSD
in the sense that the self-interaction correction is only finite for spatially
localised states, while for Bloch-like single-particle states \( E^{SIC} \)
is equal to \( E^{LSD} \). Thus, the LSD minimum is also a local minimum of
\( E^{SIC} \). A question now arises, whether there exist other competitive
minima, corresponding to a finite number of localised states, which could benefit
from the self-interaction term without loosing too much 
of the energy associated with band formation.
This is often the case for rather well localised electrons like the 3$d$
electrons in transition metal oxides or the 4\( f \) electrons in rare earth
compounds. It follows from minimisation of Eq. (\ref{eq1}) that within the SIC-LSD 
approach
such localised electrons move in a different potential than the delocalized
valence electrons which respond to the effective LSD potential. For example,
in the case of manganese, three (Mn$^{4+}$) or four (Mn$^{3+}$)
Mn $d$ electrons move in the SIC potential, while all other electrons feel 
only the effective LSD potential. Thus, by including
an explicit energy contribution for an electron to localise, the ab-initio SIC-LSD
describes both localised and delocalized electrons on an equal footing, leading
to a greatly improved description of static Coulomb correlation effects over
the LSD approximation. 

In order to make the connection between valence and localisation more explicit
it is useful to define the nominal valence as
\[
N_{val}=Z-N_{core}-N_{SIC},
\]
where $Z$ is the atomic number (25 for Mn), $N_{core}$ is the number of core 
(and semi-core) electrons (18 for Mn), and $N_{SIC}$ is the number of localised, 
i.e., self-interaction corrected, states (either three or four for 
Mn$^{4+}$ and Mn$^{3+}$ respectively). Thus, in this formulation the valence is equal to the 
integer number of electrons available for band formation. To find the valence we 
assume various atomic configurations, consisting of different numbers of localised 
states, and minimise the SIC-LSD energy functional of Eq. (\ref{eq1}) with respect 
to the number of localised electrons. The SIC-LSD formalism is governed by the 
energetics due to the fact that for each orbital the SIC differentiates between 
the energy gain due to hybridisation of an orbital with the valence bands and the 
energy gain upon its localisation. Whichever wins determines if the
orbital is part of the valence band or not and in this manner
also leads to the evaluation of the valence of elements involved.
The SIC depends on the choice of orbitals and its value
can differ substantially as a result of this. Therefore, one
has to be guided by the energetics in defining the most optimally
localised orbitals to determine the absolute energy minimum of
the SIC-LSD energy functional. The advantage of the SIC-LSD formalism is that 
for such systems as transition metal oxides or rare earth compounds the lowest 
energy solution will describe the situation where some single-electron states 
may not be Bloch-like. For Mn, these would be the Mn 3$d$ states, but 
not the O 2$p$ states, as trying to localise the latter is energetically unfavourable.

In the present work the SIC-LSD approach, has been implemented \cite{sic} 
within the linear muffin-tin-orbital (LMTO) atomic sphere approximation (ASA) band 
structure method, \cite{oka75} in the 
tight-binding representation \cite{AJ84}. 
%The electron wave functions are expanded in
%terms of the screened muffin-tin orbitals, and the minimisation of \( E^{SIC} \)
%becomes non-linear in the expansion coefficients.  

\section{Calculational details}

%As a function of Sr concentration LMO evolves from $Pnma$ space group (0\% of the Sr) to
%$R-3c$ (between 10\% and $\sim$50\% of the Sr) to $P63/mmc$ (for SMO).
We performed SIC-LSD calculations for both La$_{(1-x)}$Sr$_x$MnO$_3$ for $0.0 < x < 0.5$ and
SMO. 
The space group symmetry of the LSMO structure changes, as a function of Sr concentration,
from $Pnma$ (0\% of the Sr) to $R-3c$ (between 10\% and $\sim$50\% of the Sr) 
and $P63/mmc$ for SMO.
However, for the sake of comparison, we performed calculations using 
the same cubic crystal structure for all the
concentrations. Thus our results for LMO refer to a hypothetical cubic phase.
In order to highlight the effect of the electron doping by substituting Sr for La 
we also kept the lattice parameter constant at 7.32 atomic units (a.u.) - which 
is the average of the theoretical lattice parameters of ferromagnetic LMO and SMO.
The experimental lattice parameter, in going from LMO to SMO, changes by 1\%, 
i.e., from 7.45 to 7.38 a.u..

For the linear muffin-tin basis functions, we used
6{\it s}, 5{\it p}, 5{\it d}  partial waves for the La and 5{\it s}, 4{\it p}, 4{\it d}
for the Sr atoms,
and treated them as low-waves.\cite{Lambrechts} 
Including also 4{\it f}-basis functions on the
lanthanum, treated as intermediate waves, was of no substantial importance for the
final results. On the manganese  atoms only 4{\it s} and 3{\it d} partial waves were
treated  as low waves, whilst the 4{\it p}-waves  were treated as intermediate.
On the oxygen only 2{\it s} and 2{\it p} partial waves were treated as low-waves,
and 3{\it d}-waves as intermediate.
The atomic sphere radii were 4.0, 2.3 and 1.8 a.u. for the lanthanum and strontium, 
manganese and oxygen, respectively. These spheres were chosen to minimise 
the discontinuity in the Hartree potentials, giving 
an overlap volume of approximately 8\%. No empty spheres were used for the cubic system.
Care was taken to ensure that the results were converged with respect to both
the size of the screening cluster and the number of {\bf k} points for which
the one electron equations were solved. This was imperative to allow comparisons
to be made between different magnetic structures, which entailed
the use of different unit cells.
The screening clusters consisting of 111 atoms for the lanthanum, strontium and manganese
sites and 99 atoms for the oxygen atoms were used. The number of {\bf k} points used was
256 in the full Brillouin zone for the paramagnetic/ferromagnetic, G-type and A-type magnetic structures.

%We used SIC-LSDA, within the rigid band model and super-cell approach, to study the of 
%phase diagram of LSMO for $x$ ranging from 0.0 to 0.5. 
To perform calculations for LSMO with $0<x\le0.5$, we have utilized
the SIC-LSD within the rigid band model and supercell approach.
In the rigid band 
model, the variations in the band filling (reduction by up to 0.5 electrons) 
and lattice constant were the only 
variables depicting the change from the cubic LaMnO$_3$ to 
LSMO. Thus taking into consideration the dependence on the lattice constant, 
we could use the results from rigid band model to describe properties of
other perovskites where Sr may be replaced by Ca or Ba.
Supercells of the form La$_n$Sr$_m$Mn$_{(n+m)}$O$_{3(n+m)}$ (where $n=1,..,7$ and m=1,2,3)
were constructed to describe
charge ordering effects due to Mn$^{4+}$ 
occurring in the vicinity of Sr, and Mn$^{3+}$ present around La 
sites. \\

\section {The end members of phase diagram: LMO and SMO}

%For, 
%%%We studied the charge ordering of the end members of the phase diagram
%LaMnO$_3$ and SrMnO$_3$,
%%%. We find that 
%the SIC-LSD calculations found   
%%%results, in agreement with conventional wisdom, in 
%the Mn$^{3+}$ valence
%in LMO \cite{rik} and Mn$^{4+}$ valence in SMO.
%%%. We obtained G-type antiferromagnetic (AF-G) ordered ground state for SrMnO$_3$. It is insulating with
%%%with a band gap of 1.08 eV, twice smaller than the measured band gap \cite{saitoh} of 2.3 eV. 
We start with the application of the SIC-LSD to the end compounds of LSMO, namely 
LaMnO$_3$ and SrMnO$_3$.
Table \ref{smotab} summarizes the LSD and SIC-LSD results for SMO for three different magnetic structures,
and two Mn valence configurations. As seen in the table, the Mn$^{4+}$  configuration
in the G-type anti-ferromagnetic (AF-G) structure is the ground state solution.
The latter is insulating, with a band gap of 1.08 eV, 
twice smaller than the measured band gap \cite{saitoh} of 2.3 eV.
The ground state configuration corresponds to the localization of the three $t_{2g}$ electrons.
Localizing an extra $d$ electron, the $e_{g}$ one with the symmetry $d_{3z^2-r^2}$, 
giving rise to Mn$^{3+}$,
is unfavourable by more than 100 mRy, for all three different magnetic structures.
This energy difference between Mn$^{4+}$ and Mn$^{3+}$ 
configurations in the ferromagnetic (FM) SMO system decreases from 120 mRy to 93 mRy, when 
increasing the lattice parameter from 7.2 a.u. (corresponding 
to the theoretical pseudocubic FM ground state of SMO with Mn$^{3+}$) to 7.32 a.u.. 
On the other hand, Table \ref{smotab} also shows that only 5 mRy separate the G-type from A-type antiferromagnetic state.
This shows that the charge ordering energy in SMO is a much larger energy 
scale than the magnetic order and that its dependence on the lattice constant is small.
The charge transfer of 1.0$e$  and 0.1$e$ respectively from Mn and O  atoms
 to Sr atom hardly changes for the three different magnetic structures.
For each magnetic structure the localisation increases the Mn magnetic moment,
which, however does not change much between the different magnetic structures.
%Finally, we obtain an insulating ground state with a band gap of 1.08 eV, 
%twice smaller than the measured band gap \cite{saitoh} of 2.3 eV.

\begin{table}
\caption{\label{smotab}
Energy for different magnetic configurations for cubic SrMnO$_3$ referred to  
the ground state energy. The energy differences are per chemical unit cell
and the lattice parameter was taken to be 7.32 a.u..   
The $e_{g(3z^2-r^2)}$ orbital is written as 1$e_{g}$.
Also displayed are the magnetic moments (MM) for each of the magnetic structures.
}

\begin{tabular}{lllllll} 
%\hline
&AF-A & &AF-G & & FM & \\ 
\hline
& $\Delta E$ & MM  & $\Delta E$ & MM & $\Delta E$ & MM  \\
&  [mRy] &  [$\mu_B$] &  [mRy] &  [$\mu_B$]&  [mRy] &  [$\mu_B$] \\
\hline
LSDA                  & 115 & 2.60 & 108 & 2.49 & 120 & 2.59\\
SIC(3$t_{2g}$)        & 5   & 2.74 &  0  & 2.79 & 12  & 2.68\\
SIC(3$t_{2g}$+1$e_g$) & 114 & 3.44 & 116 & 3.47 & 105 & 3.44\\ 
%\hline
\end{tabular}
\end{table}

The FM cubic LaMnO$_3$ with the lattice parameter of 7.43 a.u. 
has the Mn$^{3+}$ configuration: 15 mRy separate 
this ground state from the Mn$^{4+}$ excited state. In 
comparison, for the Jahn-Teller distorted LaMnO$_3$ structure, 
we find 20 mRy energy difference.\cite{rik} Reducing the lattice constant by 1.5\%
to 7.32 a.u., we find that 
LaMnO$_3$ becomes nearly tetravalent and less than 5 
mRy separate the Mn$^{3+}$ ground state from the Mn$^{4+}$ excited 
state. We find the crossover between trivalent and tetravalent manganese
at a volume of elementary cell equal to 215 \AA$^3$.
This is close to the expected volume of 210 \AA$^3$ from pressure experiments
\cite{loa}.
The charge transfer from Mn and O atoms (respectively 0.9$e$ and 0.2$e$)
gives 1.5 more electrons on the La atom.
Thus the charge transfer is similar between SMO and LMO.
The calculations show that the energy scales between the Mn$^{3+}$ and Mn$^{4+}$
charge ordered state are much smaller for LMO than for SMO, specifically 20 times smaller.
Actually, this is an energy scale comparable to the energy difference between magnetic structures.
The competition between these energy scales in the case of Sr doping will be the subject 
of the next three Sections.

\section {Magnetic Structure and Charge Order in LSMO}

\begin{figure}
\includegraphics*[scale=0.5, angle=-90]{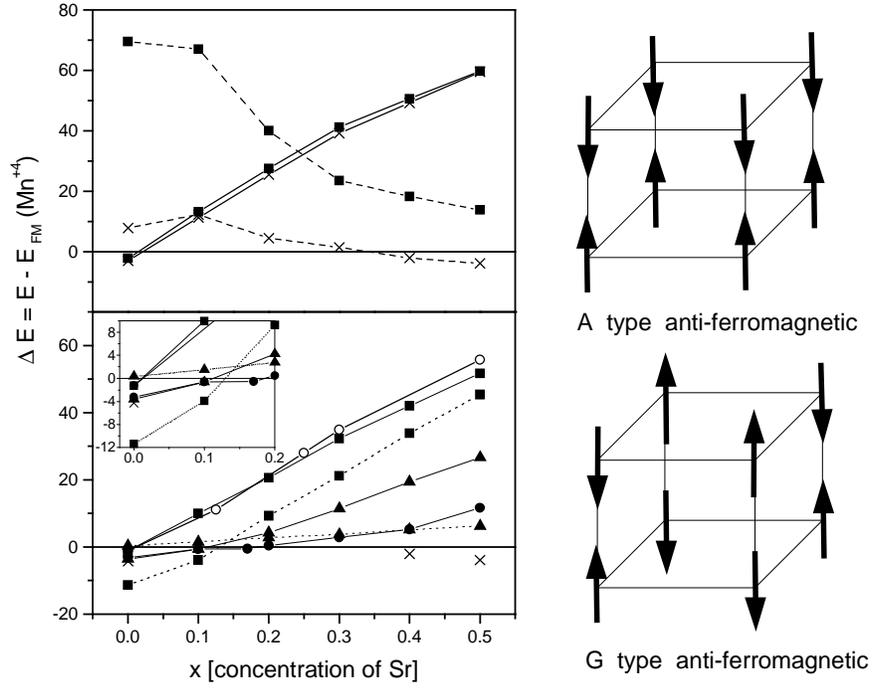}
\caption{\label{phasediag}
Left: Energy phase diagram LMO $-$ LSMO 
%(x from 0.0 to 0.5)
in the cubic structure with the lattice constant of 7.32 a.u.. 
The energy for all the systems is with respect
to the energy of FM system with Mn$^{4+}$.
Top: AF-A $-$ crosses with solid and dashed lines
for Mn$^{3+}$ and Mn$^{4+}$ respectively; AF-G $-$ 
squares with solid and dashed lines for Mn$^{3+}$ and Mn$^{4+}$, respectively.
Bottom, FM:
Mn$^{3+}$ - squares with solid line; Mn$^{4+}/$Mn$^{3+}$ supercell $-$ solid circles with solid line;
Mn$^{3+}/$Mn$^{4+}/$Mn$^{3+}$ supercell - triangles with solid line; Mn$^{4+}/$Mn$^{3+}/$Mn$^{4+}$ 
supercell $-$ triangles with dashed line; Mn$^{3+}$ for the lattice constant of 7.434 $-$ 
squares with dashed line; 
Mn$^{3+}$ for appropriate supercell $-$ open circles with solid line. 
The crosses on their own show AF-A ground state for LSMO.
Right: Schematic views of the AF-A and AF-G structures.
}
\end{figure}

In Fig.\ref{phasediag} we present the phase diagram of anti-ferromagnetically and 
ferromagneticlly ordered
LSMO for Sr concentration between 0.0 and 0.5.
For LMO, we obtain a ground state of Mn$^{3+}$ valency in an AF-A
magnetic structure. 
In the range of $x$ up to approximately 0.2 (excluding 0.0), 
the FM supercell with mixed valence of Mn$^{3+}$/Mn$^{4+}$/Mn$^{3+}$
gives the state with the lowest energy (see also inset of Fig.\ref{phasediag}).
A crossover, as a function of Sr doping, 
from a FM mixed valence Mn$^{3+}$/Mn$^{4+}$/Mn$^{3+}$ ground state to a FM Mn$^{4+}$ 
ground state occurs around 20\% Sr doping. 
For Sr concentrations between 20\% and 35\% the ground state has the valency
of Mn$^{4+}$ in a ferromagnetic structure.
For Sr concentrations larger than 40\%, the magnetic structure changes to AF-A, but 
the valency remains Mn$^{4+}$. 
For $x$ larger than 0.2, we find that the total energy scales linearly 
between Mn$^{4+}$ and  Mn$^{3+}$ configurations.
We found this by using supercells of up to 8 formula units which allowed us to model 
different distributions of  Mn$^{4+}$ and  Mn$^{3+}$ atoms.

From the top part of Fig.\ref{phasediag}, we note that for the unfavourable Mn$^{3+}$ configuration
the energy separation between the two anti-ferromagnetic structures is small.
In the Mn$^{4+}$ configuration, the energy separating these anti-ferromagnetic 
structures becomes larger. 
Obviously, three Mn band electrons are insensitive to 
the magnetic structure whilst four band electrons, including all $e_{g}$ electrons,
probe the magnetic structure.
From the bottom part of Fig.\ref{phasediag}, we note
that the Mn$^{4+}$/Mn$^{3+}$/Mn$^{4+}$ and the Mn$^{4+}$/Mn$^{3+}$ configurations
are close in energy and also close to the ground state. Increasing the Mn$^{3+}$
concentration as realized in the Mn$^{3+}$/Mn$^{4+}$/Mn$^{3+}$ configuration takes
the total energy further away from the ground state energy as the hole doping increases
through the increase of Sr concentration. Also for a slightly larger lattice constant
(7.43 a.u.) the deviation from the ground state energy steadily increases upon hole 
doping. The Mn$^{3+}$ configuration is energetically the most unfavourable. The
differences between supercell description and rigid band are small on the energy scale
of the separation to the Mn$^{4+}$ ground state and the larger the hole doping becomes the
more unfavourable the Mn$^{3+}$ configuration turns out to be.

\begin{table}
\caption{\label{halftab}
Energy for three different magnetic configurations with reference to ground state energy for 
cubic La$_{0.5}$Sr$_{0.5}$MnO$_3$.
Energy differences are per chemical unit cell and the lattice parameter was
taken to be  7.32 a.u.. $\Delta E_{RBM}$ refers to the energy difference from a
rigid band model (RBM), whilst $\Delta E_{SC}$ is the energy difference obtained using a
supercell (SC) model. The symbol MM stands for magnetic moment. 
Four different valence configurations are considered: LSDA, Mn$^{4+}$ with the
three $t_{2g}$ electrons localized and the two Mn$^{3+}$ valence configurations,
localizing each of the $e_{g}$ separately.
}

\begin{tabular}{lllll} 
%\hline
 & $\Delta E_{RBM}$  & MM  & $ \Delta E_{SC}$ & MM \\
 &             [mRy] & [$\mu_B$] &  [mRy] &  [$\mu_B$]\\
\hline
AF-A   & & &                  & \\ 
\hline
LSDA                            & 113 & 2.67 & 108 & 2.77 \\
SIC(3$t_{2g}$)                  &  0  & 3.04 &  0  & 2.99 \\
SIC(3$t_{2g}$+$e_{g(3z^2-r^2)}$)&  63 & 3.51 &  63 & 3.51\\ 
SIC(3$t_{2g}$+$e_{g(x^2-y^2)}$) &  48 & 3.53 &  47 & 3.53\\
\hline
AF-G & & &                    &\\
\hline
LSDA                             & 120 & 2.62 & 122 & 2.56 \\
SIC(3$t_{2g}$)                   & 18  & 3.02 & 22  & 2.89 \\
SIC(3$t_{2g}$+$e_{g(3z^2-r^2)}$) & 64  & 3.50 & 62  & 3.50 \\
SIC(3$t_{2g}$+$e_{g(x^2-y^2)}$)  & 64  & 3.50 & 63  & 3.50 \\
\hline
FM & & & &\\
\hline
LSDA                            & 111 & 2.69 & 113 & 2.70 \\
SIC(3$t_{2g}$)                  & 4   & 2.91 & 6   & 2.86 \\
SIC(3$t_{2g}$+$e_{g(3z^2-r^2)}$)& 56  & 3.51 & 57  & 3.50 \\
SIC(3$t_{2g}$+$e_{g(x^2-y^2)}$) & 52  & 3.52 & 54  & 3.52 \\
%\hline
\end{tabular}
\end{table}

The La$_{0.5}$Sr$_{0.5}$MnO$_3$ system was studied using both
the rigid band model and LaSrMn$_2$O$_6$ supercell. From Table \ref{halftab}, we note that
in the LSD we obtain a FM state in the rigid band model, whilst an AF-A state is obtained
in the supercell approximation. The energy differences between these magnetic structures
are 2 and 5 mRy for the rigid band model and supercell, respectively.
The SIC-LSD calculations give the same ground state both in the rigid band model and
supercell. Similarly to LSD, small energy differences of 4 and 6 mRy separate
the AF-A ground state from the FM state in the rigid band model and
supercell respectively. 
%Finally, we observe that for Mn$^{3+}$ the difference between AF-A and AF-G 
%configurations 
%is small and for Mn$^{4+}$ the difference between AF-A and AF-G
%configurations is larger.
Figure \ref{phasediag} also confirms that the rigid band model is an adequate description of the 
disordered La/Sr system. A comparison is made with supercell calculations of 
La$_n$SrMn$_{(n+1)}$O$_{3(n+1)}$ with $n=1,..,7$.
From energy differences between Mn$^{4+}$ and Mn$^{3+}$ systems for each 
$n$ we deduce an Mn$^{3+}$ ground state configuration for $n$ 
larger than 9 (for such system as La$_{10}$SrMn$_{11}$O$_{33}$) which is in agreement with $x<0.1$
from the rigid band model. This shows the close agreement between these calculations.

From Table \ref{halftab}, we also note that the energy differences between Mn$^{4+}$ and Mn$^{3+}$ 
are more or less constant with respect to the magnetic structure and also with respect to the
symmetry of the localized $e_{g}$ electron in the Mn$^{3+}$ configuration. This energy
is approximately 60 mRy. Again the energy differences between magnetic structures
are an order of magnitude lower. 

The energies in Fig. \ref{phasediag}, are closely balanced and the value of the lattice constant
will therefore be important. Increasing the lattice constant by 1\%
to 7.39 a.u., we find that for LMO the ground state remains AF-A with Mn$^{3+}$
%but for $x$ up 10\% the ground state becomes FM Mn$^{4+}$. The Mn$^{3+}$ configuration has become
%slightly less unfavourable. 
for Sr concentration $x$ up to $0.1$, and turns into FM with Mn$^{4+}$ configuration 
for $x$ larger than $0.1$. The Mn$^{3+}$ configuration has at the same time become
slightly less unfavourable. 
This we note, for example, by comparing the energy difference for
La$_{0.7}$Sr$_{0.3}$MnO$_3$ between FM Mn$^{4+}$ and Mn$^{3+}$ configurations which reduces from 32 mRy 
at lattice constant 7.32 a.u. to 21 mRy at the 1\% increased lattice constant of 7.39 a.u.. 
For La$_{0.5}$Sr$_{0.5}$MnO$_3$
even the ground state changes from AF-A at the lattice constant of 7.32 a.u. to FM at 7.39 a.u.,
the energy differences being 4 mRy and -2 mRy respectively.
This shows that the combined effect of band filling and the value of the lattice constant
gives rise to a rich variation in magnetic structure and charge order.

\begin{figure}
\includegraphics*[scale=0.5]{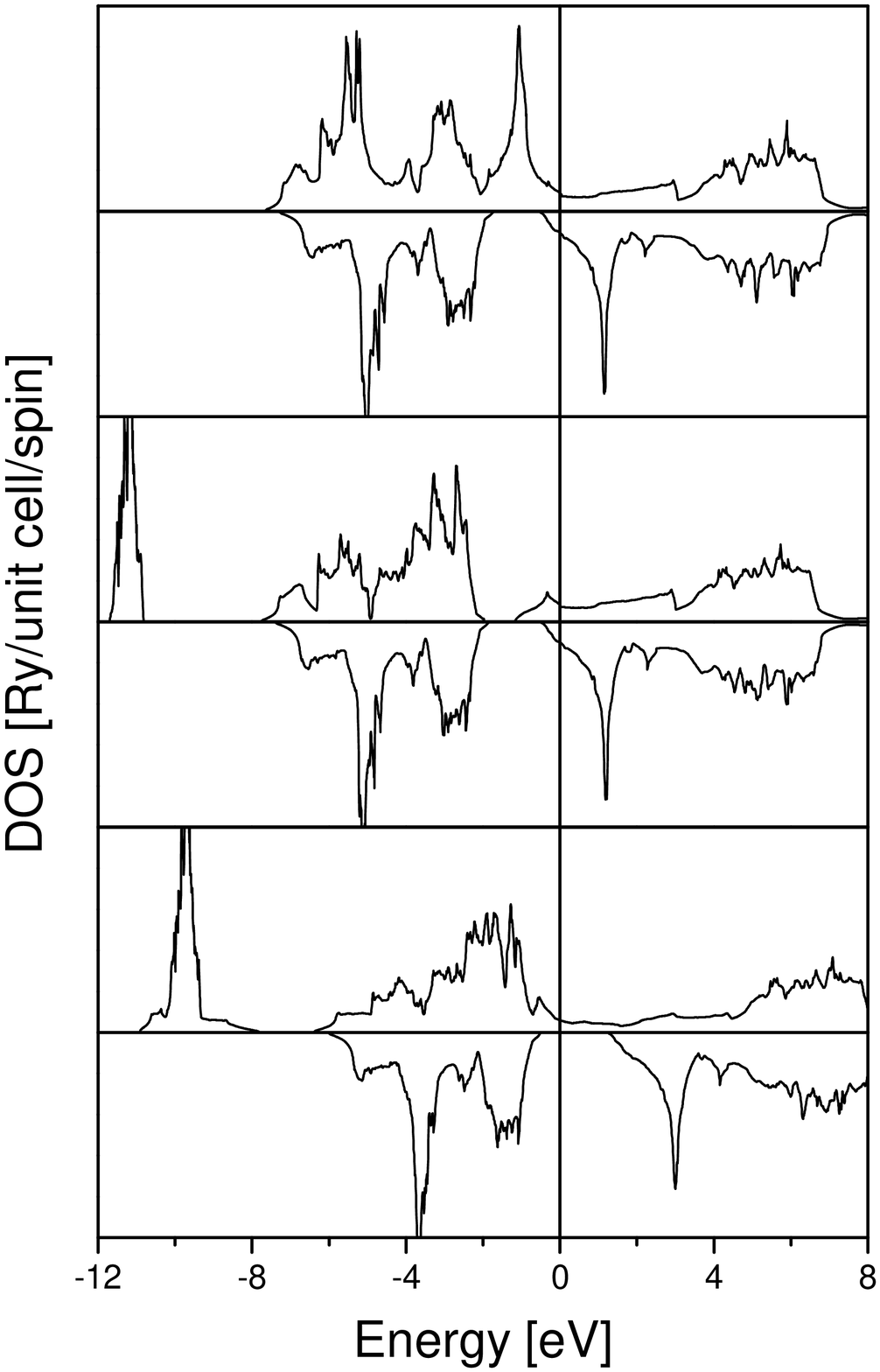}
\caption{\label{olddos}
Density of states for FM La$_{.7}$Sr$_{.3}$MnO$_{3}$ from rigid band model for 
minority and majority spin channels with respect to the Fermi energy. 
Displayed are: LSD calculation (top), SIC$-$LSD calculation for the ground state of 
three localised $t_{2g}$ electrons (centre)
and SIC$-$LSD calculation for four localised electrons (3$t_{2g}$+$e_{g(3z^2-r^2)}$) (bottom).
}
\end{figure}

The spin magnetic moments increase only slightly in comparison with LSD results.
For La$_{0.7}$Sr$_{0.3}$MnO$_3$ the  
Mn spin magnetic moment is 3.14 $\mu_B$ 
in SIC-LSD and 3.03 $\mu_B$ in LSD and 
for La$_{0.5}$Sr$_{0.5}$MnO$_3$ these are 2.91 $\mu_B$ and 2.69 $\mu_B$ respectively.
%total spin magnetic 
%moments are 3.34 $\mu_B$ in SIC-LSD and 3.31 $\mu_B$ in LSD.
The density of states for LSD calculations and the SIC-LSD in Mn$^{4+}$ and Mn$^{3+}$
configurations are shown in Fig. \ref{olddos}. 
We clearly see the majority Mn $t_{2g}$ peak, which occurs
in the LSD calculation just below the Fermi level,  
which moves down in energy below the bottom of the
valence band for the calculations for Mn$^{4+}$ and Mn$^{3+}$ valencies.
In the Mn$^{4+}$ ground state configuration
for La$_{0.7}$Sr$_{0.3}$MnO$_3$ a metallic state is obtained and the electronic 
structure in the vicinity of the Fermi level is similar to the LSD, i.e.,
it is also a nearly half-metallic system. However, for the Mn$^{3+}$ configuration, we obtain  
a half-metal with 1.6 eV band gap in the majority spin channel.
We will see in the following that the occurence of a half-metallic
density of states is closely associated with the Mn$^{3+}$ valency.

The importance of local effects was studied with supercells. From calculations for LaSrMn$_2$O$_6$
we note that the O atom in SrO layer (O$_{Sr}$) lost twice as many electrons as the O
atom in the LaO layer (O$_{La}$).
This leads to an increased spin moment of 0.2 $\mu_B$ for O$_{Sr}$ in comparison with 0.07 $\mu_B$
for O$_{La}$.
From calculations for La$_2$SrMn$_3$O$_9$, we find that Mn sandwiched between two LaO layers
loses 0.07 electrons more than Mn between a LaO and SrO layer. However, this increases the 
magnetic moment for Mn sandwiched between LaO by 0.04 $\mu_B$ only. A similar 
effect has been noted for calcium perovskites \cite{picket96}.

\section  {Oxygen Displacements and Charge Order in LSMO}

\begin{figure}
\includegraphics*[scale=0.5, angle=-90]{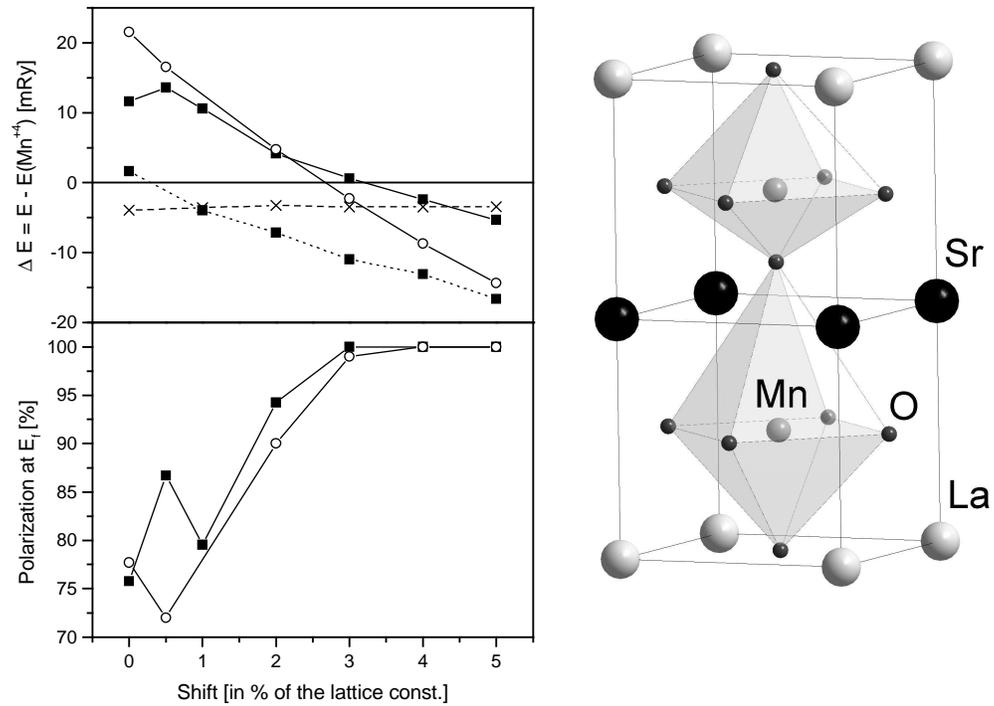}
\caption{\label{shift}
Left: In the top panel the energy difference for the FM system in the Mn$^{4+}$/Mn$^{3+}$ mixed valence state: 
%On the top energy of FM systems with mixed valence 
%Mn$^{4+}$/Mn$^{3+}$ unit cell:
squares with solid line - rigid band model (for x=0.5);
open circles - supercell model (LaSrMn$_{2}$O$_{6}$);
squares with dotted line - rigid band model (for x=0.3), and 
AF-A with Mn$^{4+}$ - crosses with dashed line. 
%All energies referenced to energy of FM with Mn$^{4+}$.
All energies are with respect to the energy of the FM system with Mn$^{4+}$.
In the bottom panel: polarisation of electrons at the Fermi level:
squares - from rigid band model (for x=0.5) and 
open circles for supercell model (LaSrMn$_{2}$O$_{6}$).
Right: Schematic view of the unit cell.
}

\end{figure}

We investigated the influence, on the Mn valency, of a tetragonal shift of the oxygen atom 
away from the SrO layer.
An upward shift in the $z$-direction of 5\% of the lattice constant was implemented.
We calculated the total energies for a double unit cell, in the FM state,
of La$_{(1-x)}$Sr$_{x}$MnO$_3$
(with $x$=0.3 and 0.5) of the following configurations: all Mn$^{4+}$, all Mn$^{3+}$, and
mixed Mn$^{4+}$/Mn$^{3+}$ (Fig. \ref{shift}).
For the mixed valence configuration, the Mn$^{3+}$ atom is taken to be 
inside the octahedron which is elongated by the tetragonal shift
(see the cell on the right hand side of Fig. \ref{shift}).
In the case of  La$_{0.5}$Sr$_{0.5}$MnO$_3$, we found that 
increasing the tetragonal shift for the oxygen atom decreases the difference  between 
the energy of the ground state (AF-A with Mn$^{4+}$) and the energy of the mixed 
valence Mn$^{3+}$/Mn$^{4+}$ configuration. 

We have calculated a change in the AF Mn$^{4+}$ ground state configuration as a function of
tetragonal shift of O atom. For La$_{0.5}$Sr$_{0.5}$MnO$_3$ the new mixed valence Mn$^{4+}$/Mn$^{3+}$
ground state occurs at 4\% shift in the rigid band model and a 3\% shift in the supercell model.
In both models the Mn$^{3+}$ configured atom was placed in the elongated MnO$_6$ octahedron.
Placing the Mn$^{4+}$ in the elongated octahedron and the Mn$^{3+}$ in the squashed octahedron 
was the most unfavourable scenario. The crossover to a new ground state also depends on the Sr 
concentration and for La$_{0.7}$Sr$_{0.3}$MnO$_3$ we find the new ground state already at 1\%
upward shift of O.

The difference between 
rigid band model and supercell results for La$_{0.5}$Sr$_{0.5}$MnO$_3$
shows the important role of the local environment, such as
presence of Sr atoms, for the electronic properties of this material.
Also, the magnetic moment on Mn$^{3+}$ increases slightly from 3.52 $\mu_B$ to 3.63 $\mu_B$, 
which is connected with 
0.1$e$ larger charge transfer from this atom, and decreases from 2.91 $\mu_B$ to 2.73 $\mu_B$ on Mn$^{4+}$.
The oxygen atom, shifted in the supercell calculation, has a magnetic moment of 0.22 $\mu_B$, twice 
larger than in the undistorted structure.

\begin{figure}
\includegraphics*[scale=0.5]{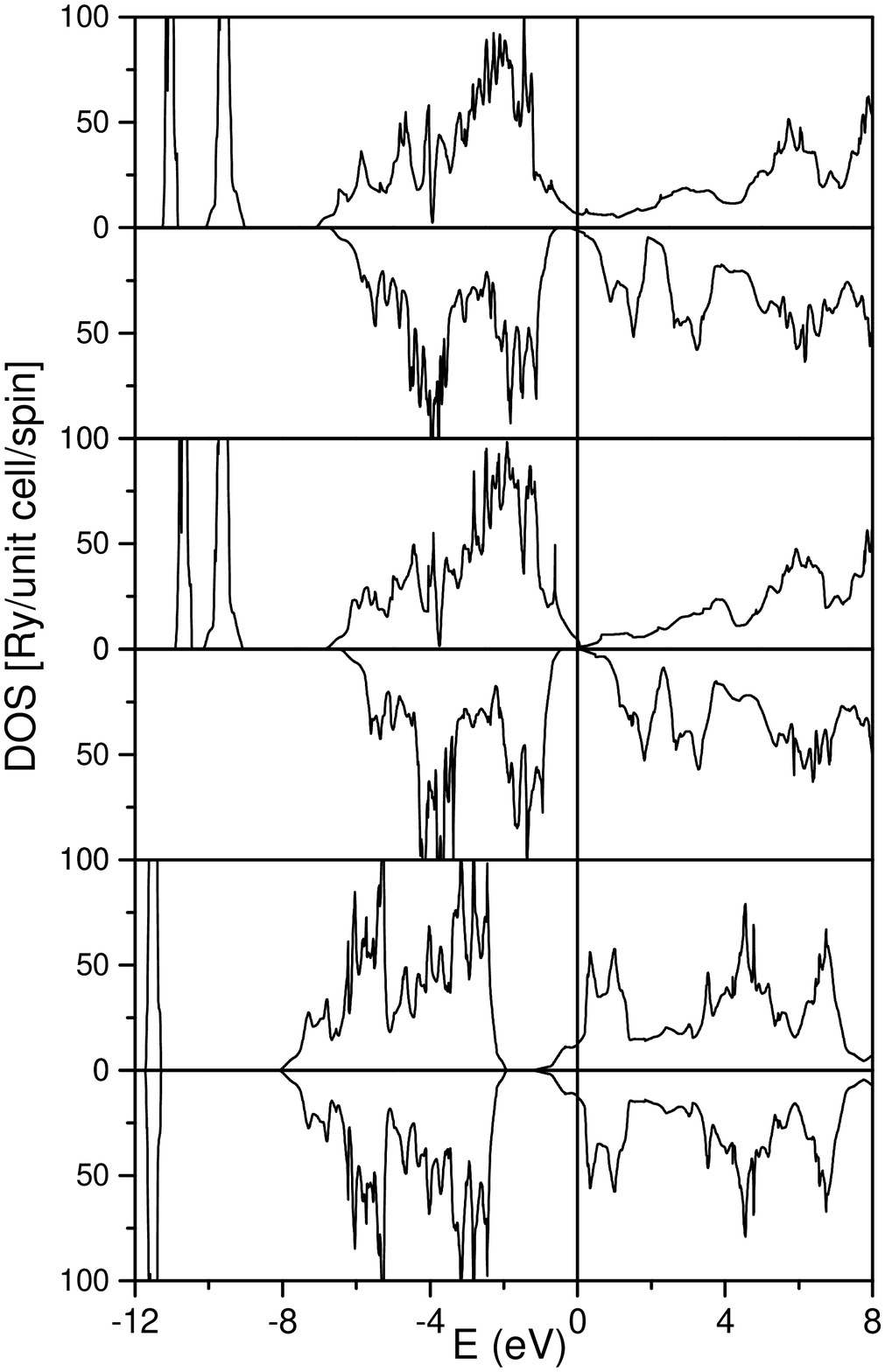}
\caption{\label{shiftdos}
DOS for LaSrMn$_{2}$O$_{6}$ with mixed valence Mn$^{4+}$/Mn$^{3+}$  
configurations for minority and majority spin channel:
is metallic without tetragonal shift for oxygen atom (top),
is half-metallic with a tetragonal shift applied to the oxygen atom of 0.04 of the lattice constant 
(centre).
In the bottom panel is the DOS for the antiferromagnetic ground state of Mn$^{4+}$ 
without tetragonal shift.
}
\end{figure}

In the bottom panel of Fig. \ref{shift} we present the polarisations of the electrons at the Fermi level. 
For shifts larger than 3\% of the lattice constant 
the system becomes fully polarised at the Fermi level and we obtain a half-metallic state. 
In Fig. \ref{shiftdos} we show the density of states for 
mixed valence Mn$^{4+}$/Mn$^{3+}$ system without tetragonal shift, mixed valence 
Mn$^{4+}$/Mn$^{3+}$ system with tetragonal 
shift for oxygen atom (4\% of lattice constant)
and for AF-A system with Mn$^{4+}$ configuration.
When increasing the shift, the bottom of the conduction 
band moves up in energy to 0.07 eV above the Fermi level. 
This system then becomes half-metallic with a band gap in one spin-channel equal to 0.48 eV.

\begin{figure}
\includegraphics*[scale=0.5, angle=-90]{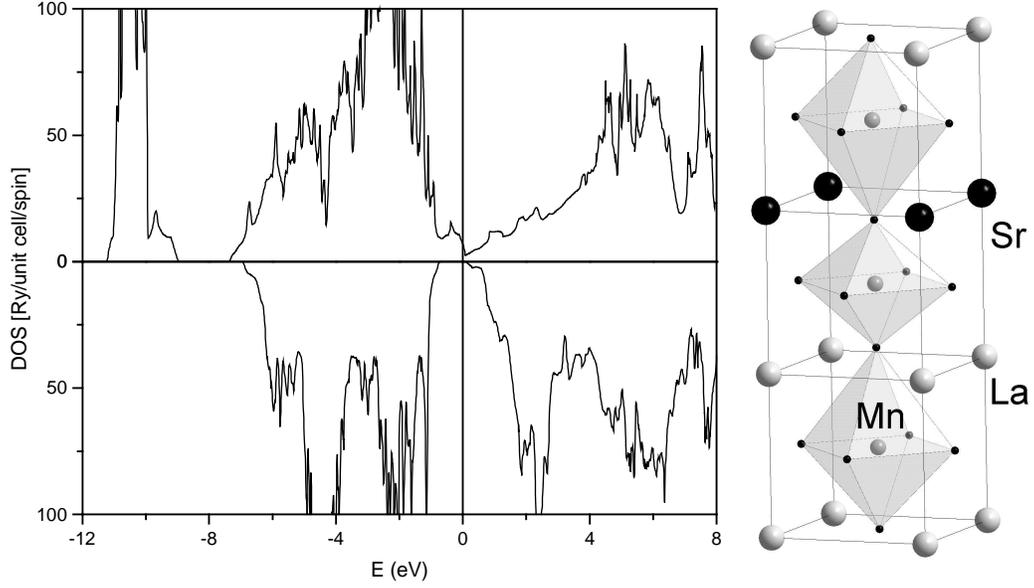}
\caption{\label{hmlsmo}
Left: DOS, in minority and majority spin channels, for La$_{2}$SrMn$_3$O$_9$ with symmetrically shifted 
(by 0.05 of lattice constant) oxygen atoms for the mixed Mn$^{3+}$/Mn$^{4+}$/Mn$^{3+}$ configuration. 
%Mn$^{3+}$/Mn$^{4+}$/Mn$^{3+}$ of manganese atoms.
Right: Schematic view of the unit cell.
}
\end{figure}

In Fig. \ref{hmlsmo} we present a different tetragonal displacement involving two
symmetrical shifts of 5\% of the lattice parameter for two oxygen atoms from LaO
and SrO layers in La$_{2}$SrMn$_3$O$_9$. For the mixed valence ground state configuration 
Mn$^{3+}$/Mn$^{4+}$/Mn$^{3+}$ we obtain a half-metallic state. In comparison with 
the above Mn$^{4+}$/Mn$^{3+}$ double unit cell, the band gap has become nearly
twice larger, at 0.8 eV.
This is due to the higher concentration of Mn$^{3+}$ atoms in the latter supercell.
The results of these two implementations of tetragonal distortions show that 
if random Jahn-Teller distortions occur,\cite{despina} the system can become
locally half-metallic.

\section  {Phase separation and Charge Order in LSMO}

Supercells in the FM regime were also constructed to model both the influence 
of Mn$^{4+}$ and Mn$^{3+}$ ordering on the total energy and the influence 
of the ordering 
of the LaO and SrO planes on the Mn charge order. Specifically, 
supercells of the form La$_n$SrMn$_{(n+1)}$O$_{3(n+1)}$ with $n=2,..,7$ and 
La$_4$Sr$_2$Mn$_{6}$O$_{18}$, and La$_4$Sr$_4$Mn$_8$O$_{24}$ were studied. 
Calculations for these systems were in agreement with the results of Fig. \ref{phasediag}.
Specifically, for $n$ smaller than $4$, the La$_n$SrMn$_{(n+1)}$O$_{3(n+1)}$ systems 
had an Mn$^{4+}$ ground state and for $n$ larger than 4, the mixed
valence Mn$^{4+}$/Mn$^{3+}$ system became the ground state.
%in agreement with the results of Fig. \ref{phasediag}. 

%xxxxxxxxxNOW DISCUSS THE NEW TABLExxxxxxxxxxxxxxxxxxxxxxxxxxxxxxxxxxxxxxxx
For the mixed valence configuration, we investigated the influence of different distributions 
of Mn$^{4+}$ and Mn$^{3+}$ on the total energy. In particular, 
we studied different scenarios of distributions of Mn$^{4+}$/Mn$^{3+}$ atoms from
all Mn$^{4+}$ (denoted by (444444) in Table \ref{disttab}) to all Mn$^{3+}$ 
(referred to as (333333) in Table \ref{disttab}), using both a six chemical units rigid band model of  
La$_{0.83}$Sr$_{0.17}$MnO$_{3}$ and the La$_5$SrMn$_{6}$O$_{18}$ supercell. 
In the supercell, the SrO layer was taken to be at the bottom of the cell.
%(at the beginning of every scenario in Table \ref{disttab}).
The energy differences are small and to emphasize this we write them down in meV.
They are nearly equal to the magnitude of magnetic structure energy differences.
The distribution 434343 of Mn$^{4+}$ and Mn$^{3+}$ valencies was the most
favourable energy state at 17\% Sr concentration for both the supercell and the
rigid band model calculations. A different distribution of the same amount of
Mn$^{4+}$ and  Mn$^{3+}$ atoms denoted by 444333 is about as unfavourable
as localizing an extra Mn $d$ electron, as realized in 334433 scenario, 
or delocalizing an Mn $d$ electron
as it happens in 443344 scenario. The energy differences between rigid band model
calculations and the supercell show that SrO and LaO layers do matter. 
In the rigid band model we see smaller energy differences upon increasing 
the amount of Mn$^{4+}$ cations (only 7 meV separate the 444444 state from
the ground state) than upon decreasing the amount of Mn$^{4+}$ cations
(215 meV separate the (333333) state from the ground state).
In the supercell approach this trend is not as pronounced and whilst
213 meV separate the 333333 state from the ground state, the 
444444 configuration is further away from the ground state, by 27 meV, than 
in the rigid band model.

%xxxxxxxxxxxxxxxxxxxxxxxxxxxxxxxxxxxxxxxxxxxxxxxxx

\begin{table}
\caption{\label{disttab}
Total energy differences (in meV), with respect to the state with the lowest
energy, for different distributions of manganese ions, 
in six units supercell in the rigid band  
(corresponding to La$_{0.83}$Sr$_{0.17}$MnO$_3$) 
and, in the supercell La$_5$SrMn$_6$O$_{18}$. 
%The scenario correspond to the ionics order 
%Mn$^{4+}$ (4) and  Mn$^{3+}$ (3) in supercell. 
For La$_5$SrMn$_6$O$_{18}$ 
the system is ordered as 
SrO-LaO-LaO-LaO-LaO-LaO and the Mn$^{4+}$ or Mn$^{3+}$ ions are sandwiched in between.
}

\begin{tabular}{lll} 
%\hline
Scenario & Supercell  & Rigid Band Model  \\
% &[meV] &  [meV] \\
%\hline
\hline
444444 & 27 & 7  \\
\hline
444443 & 36 & -- \\
444434 & 34 & 7  \\
444344 & 48 & -- \\
\hline
443344 & 54 & 41 \\
444343 & 27 & 11 \\
434434 & 45 & 18   \\
\hline
434343 & 0 &  0  \\
444333 & 79 & 63 \\
\hline
334433 & 104 & 102 \\
334343 & 59 & 52 \\
343343 & 45 & 43 \\
\hline
333333 & 213 & 215 \\
%\hline
\end{tabular}
\end{table}

The La$_4$Sr$_2$Mn$_{6}$O$_{18}$ system, i.e. increasing the Sr concentration to $\sim$30\%, 
acquires the Mn$^{4+}$ ground state configuration and is metallic. 
For this system we have studied the influence of the distribution of 
SrO and LaO layers on the total energy. The state with the lowest
energy was attained by phase separating the SrO layers from the LaO
layers, namely the two SrO layers were nearest neighbors. 
Separating the two SrO layers by one
and two LaO layers, respectively, increases respectively the energy of the system by 2 and 5 mRy.

\begin{figure}
\includegraphics*[scale=0.5, angle=-90]{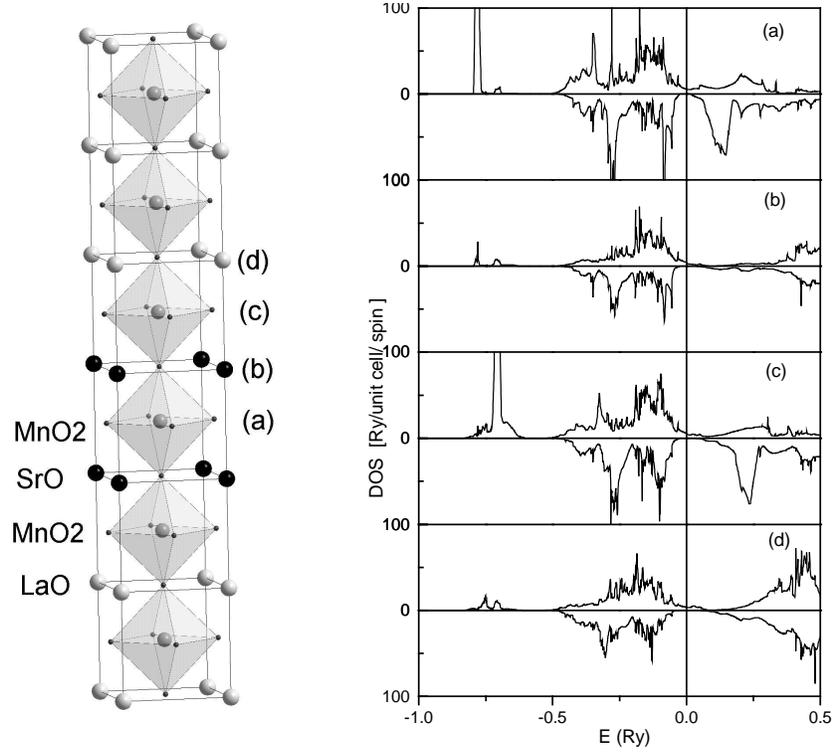}
\caption{\label{fig}
Local DOS for a supercell La$_4$Sr$_2$Mn$_6$O$_{18}$ (right). 
The configuration is (a) one MnO$_2$ layer of Mn$^{4+}$ 
sandwiched between the SrO layers (marked as black balls) 
and Mn$^{3+}$ in all other MnO$_2$ planes. Here (b) refers to SrO 
plane, (c) to MnO$_2$ plane, and (d) to LaO plane. The left-hand-side picture shows the structure.
}
\end{figure}

In Fig. \ref{fig}, we show density of states (DOS) of one of the configurations of  
La$_4$Sr$_2$Mn$_6$O$_{18}$. 
The configuration is such that we have one MnO$_2$ plane with Mn$^{4+}$,
sandwiched between two SrO planes, and all other MnO$_2$ 
planes are occupied by Mn$^{3+}$. This supercell can be 
considered as a model of phase separated LSMO which was  studied by Koida.\cite{koida} 
It is also of relevance to the La$_{0.7}$Ca$_{0.3}$MnO$_3$ system 
investigated by Bibes.\cite{bibes} 
As can be seen from Fig. \ref{fig}, this system LSMO is half-metallic. Energetically 
unfavourable, however, the energy difference between this 
state and the ground state is reduced to 20 mRy, in 
comparison with the energy difference between all Mn having 
the Mn$^{3+}$ configuration and the ground state (33 mRy). 
The gap is reduced from 0.68 eV (the scenario with the 
configuration of Mn$^{3+}$ for all manganese atoms) to 0.54 eV. 
Note that this pseudogap is constant for all MnO$_2$ layers, 
independently of their valence. None of the supercells had 
this Mn$^{3+}$ valence in their ground state configuration. However 
one could speculate that maybe at the surface Mn$^{3+}$ could 
be the stable configuration since the lower coordination 
would favour the more localised state. More complex supercells, where more Sr 
layers could phase separate, could plausibly lead to a half-metallic ground state.

The magnetic moment increases from 2.82 $\mu_B$ for Mn sandwiched between two SrO planes 
(denoted by Mn$_{SrO-SrO})$ to 2.99$\mu_B$ 
for Mn$_{LaO-LaO}$ (Mn sandwiched between two LaO planes) in the Mn$^{4+}$ configured ground state and
from 3.45$\mu_B$ for Mn$_{SrO-SrO}$ to 3.67$\mu_B$ for Mn$_{LaO-LaO}$
in the half-metallic Mn$^{3+}$ configuration.
We note that the magnetic moments on the oxygen atoms in the SrO layers are dramatically 
reduced in the vicinity of Mn$^{3+}$, namely, a value of 0.01$\mu_B$ is obtained, and they have values of 
0.15 and 0.2$\mu_B$ in the vicinity of Mn$^{3+}$/Mn$^{4+}$ and Mn$^{4+}$ respectively. 
The magnetic moments of the O atoms residing in the LaO plane remain
small at 0.05$\mu_B$ independently of the Mn valency.
In contrast, the O magnetic moments in the MnO$_{2}$ planes are small with values
around 0.05$\mu_B$ and do not change much depending on Mn valency
except for Mn$^{4+}$ at the SrO/LaO interface were an O magnetic moment as
large as 0.1$\mu_B$ has been calculated.

%xxxxxxxxxxNOW DISCUSS La$_3$Sr$_3$Mn$_6$O$_{18}$ . xxxxxxxxxxxxx
Through the study of the La$_4$Sr$_4$Mn$_8$O$_{24}$ supercell we model an even further 
increase of the Sr concentration to 50\%.
The geometry of the supercell is four La layers separated from the four Sr layers.
%From the study of La$_4$Sr$_4$Mn$_8$O$_{24}$ supercell, in which the four La
%layers are separated from the four Sr layers, 
For this supercell the ground state configuration is all Mn$^{4+}$ and this has
31 mRy lower energy than the half-metallic system with mixed valence Mn$^{4+}$/Mn$^{3+}$ configuration
(33333444) and 
61 mRy lower than the configuration with all Mn$^{3+}$. 
The magnetic moments, which are smaller than for 30\% Sr, are increasing, as is also the case for 30\% Sr, 
from 2.71 $\mu_B$ for Mn$_{SrO-SrO}$ 
to 3.05 $\mu_B$ 
for Mn$_{LaO-LaO}$ in the Mn$^{4+}$ configured ground state, and
from 3.56$\mu_B$ for Mn$_{SrO-SrO}$ to 3.61$\mu_B$ for Mn$_{LaO-LaO}$
in the half-metallic Mn$^{3+}$ configuration.
For the latter, more localised configuration, 
we note a large sensitivity of the magnetic moments of the oxygen atoms 
from the LaO layers to the presence of the SrO/LaO interface. 
At the interface the O magnetic moment of the LaO layer practically disappears. Away
from the interface, this moment is 0.1 $\mu_B$ and remains constant in these LaO layers.
In the SrO layers the O magnetic moment is slightly larger, at 0.13$\mu_B$, and remains 
constant in all SrO layers including the interface one. 
%For the polarisation of Mn atoms. 
%Decreasing ionisation of manganese atoms to Mn$^{3+}$ at interface and LMO issue disapear
%magnetic moment on oxygen atom in near to interface LaO layer and increasing from 0.03 $\mu_B$ to 
%0.1 $\mu_B$ in next layer of LaO.
%Finally we observed fluctuations of magnetic moments on oxygen atoms depend on closeness
%of interface to SMO or single SrO layer.

%xxxxxxxxxxxxxxxxxxxxxxxxxxxxxxxxxxxxxxxxxxxxxxxxxxxxx

\section {Conclusions}

We have found that the ground state configuration of LSMO consists of manganese in the
${4+}$ valence configuration. For Sr concentrations less than 20\%, a mixed valence Mn$^{4+}$/Mn$^{3+}$
ground state was obtained, whilst for concentrations in excess of 20\% a Mn$^{4+}$ ground 
state was obtained. Also, a close competition between FM and AF-A Mn$^{4+}$ states was
seen with a crossover from FM to AF-A Mn$^{4+}$ around 35\% Sr concentration. The Mn$^{4+}$ 
ground state has marginal spin polarization at the Fermi level. We found, however, that
the Mn$^{3+}$ valency was not that energetically unfavourable and that small increases
in the lattice parameter would lead to the occurence of Mn$^{3+}$ in a mixed valence 
Mn$^{4+}$/Mn$^{3+}$ state. Likewise, elongations of the MnO$_{6}$ octahedra lead
to the formation of Mn$^{3+}$ valence atoms. With the appearance of Mn$^{3+}$ ions
a half-metallic state is obtained. Furthermore, in the mixed valence state we found
that the total energy is minimized by an ordered array of Mn$^{4+}$ and Mn$^{3+}$ MnO$_{2}$  
planes. The relation of this finding to the observed charge-ordered stripes\cite{peterl}
will be the topic of a further study.

As the Sr doping increases, the Mn$^{4+}$ state becomes more and more favoured with respect
to the Mn$^{3+}$. The Sr doping results in one more valence band electron which is obtained
from delocalizing a Mn$^{3+}$ $e_{g}$ state. In other words, Sr hole doping favours 
band formation instead of localization.

Finally, for La$_{0.7}$Sr$_{0.3}$MnO$_{3}$, of huge relevance to spintronics
applications, we suggest the following scenario to explain the half-metallic
properties. 
%The  bondlengths of Mn-O in distorted LMO
%change from 1.907 $\AA $ to 2.179 $\AA $, depending on direction,
%whilst in La$_{0.7}$Sr$_{0.3}$MnO$_3$, from x-ray results, every
%Mn-O bondlength is the same at 1.949 $\AA$.\cite{alatlsmo} 
Whilst in La$_{0.7}$Sr$_{0.3}$MnO$_3$, from x-ray results, every
Mn-O bondlength is the same at 1.949 $\AA$,\cite{alatlsmo}
the  bondlengths of Mn-O in distorted LMO
change from 1.907 $\AA $ to 2.179 $\AA $, depending on direction.
It seems however, based on pulsed-neutron diffraction measurements,\cite{despina}
that the Jahn-Teller distortion is still present in LSMO and could
therefore, according to our analysis, lead to the formation of Mn$^{3+}$ ions, which would
be sufficient to give rise to a half-metallic state.

\section*{Acknowledgements}
G.B. was supported by the EU-funded Research Training  Network: "Computational Magnetoelectronics" (HPRN-CT-2000-00143).
%The authors thank Drs. A. Haznar and R. Tyer for useful comments.  
 
%

\newpage

\end{document}